\documentclass[12pt,manuscript]{emulateapj}

\newcommand{\beq}{\begin{equation}}
\newcommand{\eeq}{\end{equation}}
\newcommand{\bea}{\begin{array}}
\newcommand{\eea}{\end{array}}

\shorttitle{Habitable Planets in OGLE-06-109L}
\shortauthors{Wang and Zhou}

\begin{document}

\title{Forming Habitable Planets around Dwarf Stars: Application to OGLE-06-109L}

\author{Su Wang and Ji-Lin Zhou}
\affil{Department of Astronomy \& Key Laboratory of Modern Astronomy
and Astrophysics in Ministry of Education, Nanjing University,
Nanjing 210093, China, suwang@nju.edu.cn, zhoujl@nju.edu.cn}

\begin{abstract}
Dwarf stars are believed to have small protostar disk where planets
may grow up. During the planet formation stage, embryos undergoing
type I migration are expected to be stalled at inner edge of
magnetic  inactive disk ($a_{\rm crit} \sim 0.2-0.3 ~$AU). This
mechanism makes the location around $a_{\rm crit}$ a sweet spot of
forming planets. Especially, $a_{\rm crit}$ of dwarf stars with
masses $\sim 0.5~M_\odot$ is roughly inside the habitable zone of
the system. In this paper we study the formation of habitable
planets due to this mechanism with a model system OGLE-06-109L. It
has a $0.51~ M_\odot$ dwarf star with two giant planets in 2.3 and
4.6 AU observed by microlensing. We model the embryos undergoing
type I migration in the gas disk with a constant disk accretion rate
($\dot M$). Giant planets in outside orbits affect the formation of
habitable planets through secular perturbations at the early stage
and secular resonance at the later stage. We find that the existence
and the masses of the habitable planets in OGLE-06-10L system depend
on both $\dot M$ and the speed of type I migration.  If planets
formed earlier so that $\dot M$ is larger ($\sim 10^{-7} M_\odot $
yr$^{-1}$), terrestrial planet can not be survived unless the type I
migration rate is  an order of magnitude less. If planets formed
later so that $\dot M$ is smaller ($\sim 10^{-8} M_\odot $
yr$^{-1}$), single and high mass terrestrial planets with high water
contents ($\sim 5\% $) will be formed by inward migration of outer
planet cores. A slower speed migration will result in several
planets by collisions of embryos, thus their water contents are low
($\sim 2\%$). Mean motion resonances or apsidal resonances among
planets may be observed if multiple planets survived in the inner
system.
\end{abstract}

\keywords{(\textit{stars:}) planetary systems: formation -- solar
system: formation -- stars: individual (OGLE-06-109L)}

\section{Introduction}

One of the interesting problems for planetary dynamics is whether
there exist stable orbits in the habitable zone (HZ) of main
sequence stars. Earth-size planets in HZ possibly have similar
compositions especially liquid water,  which is important for life
survival (Kasting et al. 1993; Hinse et al. 2008). Currently, radial
velocity technique is not sensitive enough to detect  terrestrial
planets in HZ of $\sim 1$ AU around solar type stars. However, dwarf
stars with masses $\sim 0.5 ~M_\odot$ have very close habitable
zones ($\sim 0.2-0.3$ AU) due to their low stellar luminosity, thus
they are good candidates for terrestrial planets detection in HZ.

According to the simulations of terrestrial planet formation  in
exoplanet systems, {\em in site} formation of Earth-size planets in
close-in orbits requires a very massive circumstellar disk, e.g.,
$\sim 30$ times of minimum solar nebular (Raymond et al. 2008) to
form a $\sim 10~ M_\oplus$ planet around GJ 581. Thus formation of
Earth-like planets in close-in orbits requires either keeping adding
building blocks into inner orbits, or massive cores formed in
outside orbits and migrating inward under, e.g., the type I
migration.

If giant planets already formed in outside orbits, perturbations
from giant planets that undergo type II migrations may be helpful to
the final assembly of terrestrial planets in close-in orbits. During
the inward migration of giant planets embedded in the viscous gas
disk, the locations of inner mean motion resonances (MMR) with the gas giants , mainly
2:1 MMR, will sweep through the inner disk, trap
and shepherd the embryos, and excite their eccentricities, which
results in the mergers of isolated embryos and the formation of
close-in planets (Zhou et al. 2005; Fogg \& Nelson 2005; Raymond et
al. 2006). During the stage of gas disk  depletion, secular
resonances caused by disk potential and outer giant planets will sweep
into the inner region, which may excite the eccentricities of the
embryos and results in the merger and  formation of exoplanets
(Nagasawa et al. 2003; Zhou et al. 2005).

In the above mentioned scenarios,  the type I migration of embryos
is not considered. Due to the fast rate of type I migration, the
sweeping of 2:1 MMR (caused by type II migration of outside giant
planets) and secular resonances may be lagged the fast inward
migration of embryos, thus their effects will be greatly reduced. In
that case, the embryos will  migrate unless there are some
mechanisms to shop it. Although in Fogg \& Nelson 2007 they considered
the type I migration in hot Jupiter systems,
the formation of terrestrial planets in close-in orbits
under the perturbations of giant planets in the outer
region need to be re-investigated to see whether secular perturbations and
secular resonances are effective.

Recent hydrodynamical simulations indicate that a location
of density maximum in the gas disk can stop the type I inward
migration \cite{mas06}. During the evolution of the protoplanet,
the exchange of angular momentum between the protoplanet and the
gas disk generates a corotation torque on the protoplanet with the
expression \cite{ward91,masset01,masset02},
 $\Gamma_c \propto \Sigma_g d\ln (\Sigma_g/B)/d\ln a$,
where $\Sigma_g$ is the surface density of the gas disk, B is the Oort constant and $B  \propto a^{-3/2}$  in a near Keplerian motion.
 If the surface density can jump over $3\sim 5$ gas disk
thickness, the positive corotation torque will overcome the negative
Lindblad torque which casues the inward type I migration,
making the location of density maximum a possible
position to trap the protoplanets \cite{mas06}.

Several mechanisms can lead to the density maximum in the
gas disk, e.g., {\em the inner disk cavity}. At the place where
the toque induced by the stellar magnetic field dominates over
disk internal stresses, material in the disk will fall onto the
surface of the star thus the disk was truncated \cite{Kon91}. This
place is roughly the corotation radius of the star, with a maximum
stellar distance
 $\sim $ 9 stellar radii. Before the protostar moving to the main
sequence, the radius is 2-3 times larger. Thus the inner disk truncation of a main sequence star with mass
0.5 $M_\odot$ is about 0.065 AU to 0.085 AU. In our simulations, we
choose 0.1 AU as the boundary of the inner cavity of the gas disk.
Once embryos reach this boundary, type I migration will be stopped
effectively.

Near the {\em inner edge of the dead zone} ($a_{crit}$) is another
location that induces density maximum in the gas disk. During the
classical T Tauri star (CTTS) phase, inside $a_{crit}$, the
protostellar disk is thermally ionized, so it is totally active
zone. While outside $a_{crit}$, stellar X-rays and diffuse cosmic
rays ionize a surface layer with a thickness $\cong 100 $g cm$^{-2}$
on either side of the protostellar disk (active layer). Between the
active layers is the dead zone near the midplane \cite{BH91,gam96}.
Near the inner edge of the dead zone in the midplane, a positive density gradient is
induced \cite{KL07,KL09}. For the protostellar disk in {\em  ad hoc
} $\alpha$-prescription \cite{SS73}, the mass accretion
rate is constant across the disk, $\dot{M}_g=3\pi \alpha c_s h
\Sigma_g/ $, where $c_s$ and $h=c_s/\Omega_K$ are the sound speed of
the midplane  and the disk scale-height, respectively. As  the value of
$\alpha$ increases from dead zone ($\sim 0.006$) to active zone
($\sim 0.018$)(Sano et al. 2000), the gas density
in the midplane of the disk increases from active zone to dead zone
about two times. Then a density maximum appears in the inner edge of
the dead zone which is helpful to halt the embryos undergoing type I
migration.

Due to the density profile enhancement at the inner edge of magnetic
inactive zone ($a_{\rm crit} \sim 0.2-0.3~$AU), embryos undergoing
type I migration are expected to be stalled there, which makes the
location around $a_{\rm crit}$ an ideal place for the accumulation
of embryos. Cohesive mergers among embryos may finally form
terrestrial planets. As the location of $a_{\rm crit}$ around dwarf
stars with mass $\sim 0.5~M_\odot$ is roughly in the HZ, the forming
planets may be most possibly habitable.

In this paper we study the formation of habitable planets due to the
above mechanism with a model system OGLE-06-109L. Among the planets
detected by microlensing, OGLE-06-109L system is the first one with
observed multiple planets. It is 1490 pc away from the Sun, with a
star of $\sim 0.51~ M_\odot$ (solar mass) and two planets of
$0.71~M_J$ (Jupiter mass) and 0.27 $M_J$ in the orbits of 2.3 AU and
4.6 AU, respectively (Gaudi et al. 2008). The eccentricity of the
out planet is 0.11 deduced from observation. In the first paper
(Wang et al. 2009, hereafter WZZ09), we have studied the
eccentricity formation of the system. The planetary scattering model
or the mean-motion resonance crossing model can be account for the
eccentricity formation. According to our models, the final
eccentricity of inner planet ($e_b$) may oscillate between [0-0.06],
comparable to that of Jupiter (0.03-0.06).

According to Kasting et al. 1993, HZ is estimated to be [0.25-0.36
AU] or larger in OGLE-06-109L system. In WZZ09, we have demonstrated
orbits in center of  HZ with semimajor axis $a\in$ [0.28 AU, 0.32
AU] would be out of HZ due to the high eccentricities excited by the
secular resonance of two giant planets in the system. An additional
inner planet or outer planet would suppress the eccentricity
perturbation and greatly improve the prospects for habitability of
the system (Malhotra \& Minton 2008). Orbits in the rest region of
HZ remain in HZ up to 10 Myr integrations. Based on the results,
OGLE-06-109L is a hopeful candidate system for hosting habitable
terrestrial planets. The organizing of the paper is as follows. In
section 2 we present the model, method and initial setup of the
simulations. The simulation results are shown in Section 3. Section
4 presents our conclusions.

\section{Model}

\subsection{Disk Model}

To model the effect of density increment near $a_{\rm crit}$ in the
gas disk, we formulate the gas disk as a viscous one with constant
mass accreting to the star. Its surface density at stellar distance
\emph{a} is given as \cite{Pringle}
\begin{equation}
\Sigma_g=\frac{\dot {M}}{3\pi\nu(a)},
\end{equation}
where $\dot {M}$ is the stellar accretion rate, $\nu (a)$ is the effective
viscosity. According to the results from observation
data of the young cluster $\rho$-Oph, the star accretion rate $\dot
M$ can be estimated to (Natta 2006)
\begin {equation}
\dot M\simeq 4\times 10^{-8}(\frac{M_*}{M_\odot})^{1.8}M_\odot {\rm
yr}^{-1}.
\end {equation}
So we adopt a baseline $\dot M  = 1 \times 10^{-8} ~M_\odot {\rm
yr}^{-1}$ for OGLE-06-109L system. However, during the earlier stage
of T Tauri stage, $\dot{M}$ could be 1-2 order of magnitude larger
\cite {hartmann}.

The effective viscosity $\nu(a)=\alpha c_sh$, where $\alpha$, $c_s$,
and $h=c_s/\Omega$ refer to the efficiency factor of angular
momentum transport, sound speed at the mid plane, and the isothermal
density scale height, respectively (Shakura \& Sunyaev 1973). To
model this effect, we let $\alpha_{\rm MRI}$ and $\alpha_{\rm dead}$
denote the $\alpha$-values of the MRI active and dead regions,
respectively. The effective $\alpha$ for the disk is modeled as
(Kretke \& Lin 2007; Kretke et al. 2009)
\begin{equation}
\alpha_{\rm eff} (a)=\frac{\alpha_{\rm dead}-\alpha_{\rm MRI}}{2}
[{\rm erf}(\frac{a-a_{\rm crit}}{0.1a_{\rm crit}})+1]+\alpha_{\rm MRI},
 \label{alpeff}
\end{equation}
where ${\rm erf}$ is the error function, $0.1 a_{\rm crit}$ is
thought as the width of the transition region. In this paper, we
adopt $\alpha_{\rm MRI}=0.01,\alpha_{\rm dead}=0.001$ (Sano et al.
2000). In all the simulations, we think the embryo to be run into
the central star if its semimajor axis is less than 0.1 AU.

The location of the inner edge of the MRI dead zone, $a_{\rm crit}$,
varies with the disk temperature, kinematics and mass accretion
rate, etc. Here we adopt the expression from Kretke et al. (2009),
\begin{equation}
a_{\rm crit}=0.16 ~{\rm AU}(\frac{\dot{M}}{10^{-8}M_{\odot}{\rm yr}^{-1}})^{4/9}
(\frac{M_{*}}{M_{\odot}})^{1/3}(\frac{\alpha_{\rm MRI}}{0.02})^{-1/5}.
 \label{acrit}
\end{equation}
During the evolution of T-Tauri star and its disk, $\dot{M}$ will
decrease on average. So the location of $a_{\rm crit}$ is migrating
inward as the time proceeds.

Combing with the empirical minimum mass solar nebula (hereafter
MMSN, Hayashi 1981),  we obtain the following surface density for
the circumstellar disk (shown in Figure 1):
\begin{equation}
\Sigma_{g}=\Sigma_{0}  (\frac{a}{\rm
1AU})^{-3/2}(\frac{\alpha_{\rm eff}}{10^{-3}})^{-1}
\exp({-\frac{t}{\tau_{\rm dep}}}),
\label{sigg3}
\end{equation}
where $\Sigma_0=883 $ g~cm$ ^{-2}$ is the initial density
corresponding to that at $\sim 1$ Myr for $\dot M  = 5 \times
10^{-8} ~M_\odot {\rm yr}^{-1}$, due to disk accretion,
photoevaporation or planet formation, the gas is depleted with a
timescale $\tau_{\rm dep}\sim 3$ Myr (Haisch et al 2001), $t$ is the
evolution time.

\begin{figure}
\begin{center}
  \epsscale{1.3}
 \plotone{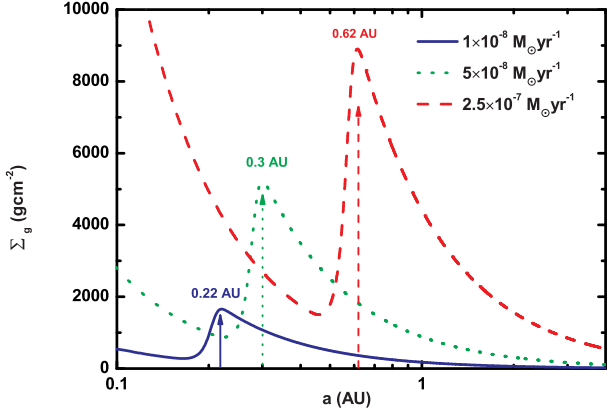}

 \vspace{0cm}

 \caption{ The density profile of the disk with different star
 accretion rate ($\dot M$). In the region outside the density
 maximum, the density profile satisfies $\Sigma_g \propto r^{-3/2}$.
 The blue solid line represents $\dot M=1\times
 10^{-8}~M_\odot yr^{-1}$, the location of the density maximum is at 0.22 AU.
 The green dot line shows the density profile of $\dot M=5\times
 10^{-8}~M_\odot yr^{-1}$, the density maximum location is at 0.3 AU.
 The red dash line means $\dot M=2.5\times
 10^{-7}~M_\odot yr^{-1}$, the location of the density maximum is at 0.62 AU.
 The inner edge of the disk is in the location of
 0.1 AU.
 \label{fig1}}
 \end{center}
\end{figure}

\subsection{type I migration and e-damping}

For a planet embedded in a geometrically thin and locally isothermal
disk, angular momentum exchanges between the planet and the gas disk
will cause a net momentum loss on the planet, which results in a
fast and so-called type I migration of the planet
\cite{GT79,Ward97,Tan02}. Some mechanisms are proposed recently to
reduce the speed or even reverse the direction of migration.
Laughlin et al. (2004), and Nelson \& Papaloizou (2004) proposed
that, in the locations where the magnetorotational instability (MRI)
is active, gravitational torques arising from  magnetohydrodynamical
turbulence will contribute a random walk component to the migratory
evolution of the planets, thus prolong the drift timescale.
Analytical estimation through Fokker Planck approach indicates the
survival rate of planet under type I and stochastic torques is
around $1-10\%$ (Adams \& Bloch 2009). Paardekooper \& Mellema
(2006) noticed that the inclusion of radiative transfer can cause a
strong reduction in the migration speed. Subsequent investigations
(Baruteau \& Masset 2008; Kley \& Crida 2008; Paardekooper \&
Papaloizou 2008) indicated that the migration process can be slowed
down or even reversed for sufficiently low mass planets. Through
full three-dimensional hydrodynamical simulations of embedded
planets in viscous, radiative disks, Kley et al. (2009) confirmed
that the migration can be directed outwards up to planet masses of
about 33 $M_\oplus$.

\begin{figure}
 \epsscale{1.3}
\plotone{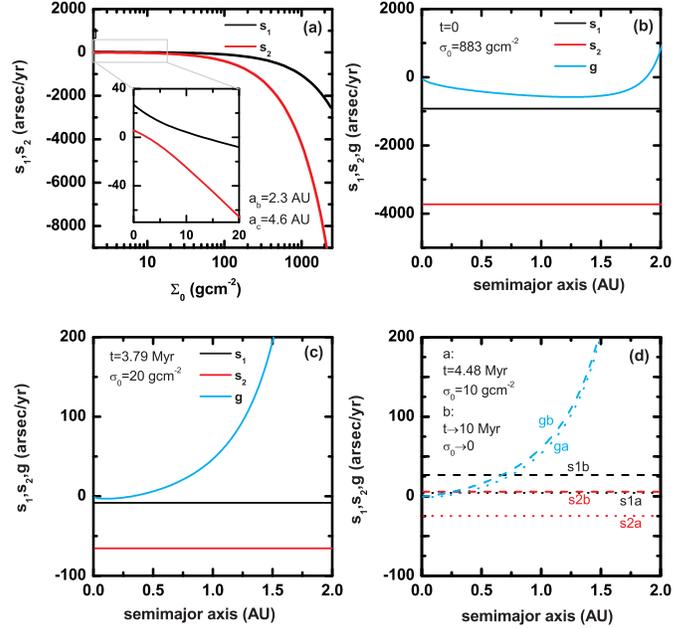}

 \caption{Location of secular resonance in the inner region of the
 system at different time during the evolution. (a) The precess
 eigenfrequencies $s_1,~s_2$ of the two giant planets due to different density
 of gas disk $\Sigma_0$ and mutual perturbation. (b) At t=0, the proper
 frequency of test particle $g$ compared to $s_1,~s_2$. There is no point of intersection between $g$ and $s_1,~s_2$.
 Secular resonance cannot happen in the inner region of the system. (c) At t=3.76 Myr, the proper
 frequency of test particle $g$ compared to $s_1,~s_2$. $g$ is close to
 $s_1$. (d) At t=4.48 Myr and 10 Myr, the proper
 frequency of test particle $g$ compared to $s_1,~s_2$. Secular
 resonance happens at the location of $\sim 0.25$ AU.
  \label{fig2}}
\end{figure}

Taken these uncertainties into considerations,  we adopt a standard
description of type I migration with a speed reduction factor $f_1$.
The timescale of type I migration for an embryo $m$ is \cite{Tan02}
\begin{eqnarray}
\tau_{migI}=\frac{a}{|\dot{a}|}~~~~~~~~~~~~~~~~~~~~~~~~~~~~~~~~~~~~~~~~~~~~~~~~~~~
\nonumber\\
=\frac{1}{f_1(2.7+1.1\beta )}
(\frac{M_*}{m})(\frac{M_*}{\Sigma_ga^2})(\frac{h}{a})^2
[\frac{1+(\frac{er}{1.3h})^5}{1-(\frac{er}{1.1h})^4}]\Omega^{-1},
\label{tauI}
\end{eqnarray}
where $r,e,~h,~\Omega,~{\rm{and}}~\Sigma_g$ are the stellar
distance, eccentricity of the embryo, scale height of the disk, the
Kepler angular velocity, and the surface density profile of gas in
the disk, respectively, $\beta=-d{\rm ln\Sigma_g}/d\rm{lna}$.
According to the density profile  in equations (\ref{alpeff}) and
(\ref{sigg3}), the factor $\beta$ will be negative when crossing the
location of density maximum to the inner disk. So when the embryo
goes across the density maximum, the embryo may be sustained in the
maximum location. In our simulations, we consider $f_1$ is in the
range of [0.001,1]. The type I migration effect is incorporated with
an additional acceleration to any embryo with mass $\le 10~M_\oplus$
(Cresswell \& Nelson 2006),
\begin{equation}
\textbf{F}_{migI}=-\frac{\textbf{v}}{2\tau_{migI}}, \label{FI}
\end{equation}
where $\textbf{v}$ is the velocity vector of \emph{m} in the
stellar-centric coordinate \ref{fig2}.

Interactions between embryos and the gas disk may damp the
eccentricities of the embryos (Goldreich \& Tremaine 1980). The
timescale of the eccentricity damping for an embryo with mass
$m<10 ~M_\oplus $ can be described as (Cresswell \& Nelson 2006),
\begin{equation}
(\frac{e}{\dot{e}})_{\rm edap}=\frac{Q_e}{0.78}(\frac{M_*}{m})
(\frac{M_*}{a^2\Sigma_g})(\frac{h}{r})^4\Omega^{-1}[1+\frac{1}{4}(e\frac{r}{h})^3],
\label{ede}
\end{equation}
where $r$,  $e$, $h$, $\Omega$ are the distance from the star,
eccentricity of the embryo, scale height of the disk and the
Kepler angular velocity, respectively, $Q_e=0.1$ is a
normalization factor to fit with hydrodynamical simulations. This
effect is included by add an extra-force term to the equations of
embory motion:
\begin{equation}
\textbf{F}_{edap}=-2\frac{(\textbf{v} \cdot
\textbf{r})\textbf{r}}{r^2\tau_{edap}}. \label{Fedp}
\end{equation}

\subsection{Initial configurations of the system  }

We assume two giant planets ($m_b,~m_c$) have already formed with
observed masses of OGLE-09-106L b, c  initially at 4 AU and 8 AU,
respectively.  The  giant planets undergo type II migration with a
fixed timescale about 1 Myr (WZZ09). An additional embryo  with mass
$16.8~ M_\oplus$ is put  between them. A planetary scatter will
occur between one giant planet and the additional embryo at about 5
Myr, just as in WZZ09, to account for the observed eccentricities of
the two giant planets.

To model the formation of planets in the terrestrial region, we put
18 embryos with isolation masses (Ida \& Lin 2004)
\begin{equation}
M_{\rm iso} = 0.12 \gamma_{\rm ice}^{3/2} f_d^{3/2} (\frac{a}{\rm
1AU})^{3/4} (\frac{M_*}{ M_\odot})^{-1/2} M_\oplus
 \label{miso}
\end{equation}
in the inner region, where $\gamma_{\rm ice}$ is the volatile
enhancement with a value of  4.2 or 1 for material exterior or
interior to the snow line (0.68 AU for OGLE-06-109L system),
respectively.
The embryos with the isolation mass are located in 10 mutual Hill radii from each other.
For a dwarf star with mass $0.51~M_\odot$, we assume a
disk with  $\rm f_d \le  2$ . For $f_d=1$, the masses of embryos
range from 0.084 $M_\oplus$ to 3.33 $M_\oplus$ at [0.39 AU, 3 AU],
with a total mass of $19.1~ M_\oplus$. For $f_d=2$, the masses of
embryos range from 0.175 $M_\oplus$ to 9.42 $M_\oplus$ at [0.26 AU,
3 AU], with a total mass of $38.5 ~M_\oplus$. The outmost embryo in
the inner region is in the orbit 3.5 Hill Radii away from $m_b$. The
giant planets and embryos are all in near-coplanar and near-circular
orbits initially (with initial eccentricities $e=10^{-3}$ and
inclinations $i=e/2$).

\subsection{Perturbation from two giant planets}

The effect of giant planets is twofold: for embryos  in nearby
orbits,  perturbations from two giant planets will excite their
eccentricities, which may be helpful for the merge of small embryos
into ones. The maximum eccentricity that an embryo initially in
circular orbits exited under the perturbation of out planet is
(Mardling 2007)
\beq e_{\rm max}= \frac{(5/2)(a_e/a_c) e_c \epsilon
^{-2}}{\left| 1-\sqrt{a_e/a_c}(m_e/m_c)\epsilon ^{-1} \right |,}
\label{emax}
\eeq
where subscribes of $e$ and $c$ denote those of
the embryo in inner orbit and the giant planet, respectively,
$\epsilon =\sqrt{1-e_c^2}$. According to equation (\ref{emax}), the
excitation of initial embryos located from [0.26, 3] AU is [0.01,
0.10] from $m_b$ and [0.007, 0.08] from $m_c$. Thus the effect of
inner giant planet is non-negligible for embryos around 3 AU.

Another effect that may contribute to the formation of planets in
habitable region is the secular resonance. Due to the presence of
gas disk and mutual perturbation, the orbits of two giant planets
will precess with eigenfrequencies ($s_1,s_2$) (Heppenheimer 1980),
\begin {equation}
s_{1,2}=\frac{1}{2}\{(A_{11}+A_{22})\pm
[(A_{11}-A_{22})^2+4A_{12}A_{21}]^{1/2}\},
\end {equation}
where
\begin {eqnarray}
\begin{array}{llll}
A_{11}=2n_ba_bm_cN_{bc}-(\pi F/4a_b)\sigma_0,
\nonumber\\
A_{12}=-2n_ba_bm_cP_{bc},
\nonumber\\
A_{22}=2n_ca_cm_bN_{bc}-(\pi F/4a_c)\sigma_0,
\nonumber\\
A_{21}=-2n_ca_cm_bP_{bc},
\end{array}
\end {eqnarray}
where $\sigma_0=\Sigma_0(\alpha_{\rm eff}/10^{-3})^{-1}
\exp(-t/\tau_{\rm dep})$. We define $a_-={\rm{min}}(a_b,a_c)$ and
$a_+={\rm {max}}(a_b,a_c)$, so that
\begin{eqnarray}
\begin{array}{ll}
N_{bc}=\frac{1}{8}(a_-/a_+^2)b_{3/2}^{(1)}(a_-/a_+),
\nonumber\\
P_{bc}=\frac{1}{8}(a_-/a_+^2)b_{3/2}^{(2)}(a_-/a_+),
\end{array}
\end{eqnarray}
where $b_{3/2}^{(i)}$ is the Laplace coefficients. $F = 4.377345$
for surface density slope $\beta=3/2$.

The proper frequency of test particle associated with the longitude
of perihelion is
\begin {equation}
g=2n_pa_p\cdot (m_bN_{pb}+m_cN_{pc})-(\pi F/4a_p)\sigma_0,
\end {equation}
where subscript $p$ refers to the test particle, terms with
$N_{pb}$, $N_{pc}$ are due to the perturbation of two planets, term
with $\sigma_0$ is due to the gas disk. The definition of $N_{pb}$
and $N_{pc}$ are similar to $N_{bc}$. When the proper frequency of
an embryo meets with either of the two eigenfrequencies, i.e.,
$g=s_i$, (i=1, 2), secular resonance occurs, which may excite the
eccentricities of embryos.

The two eigenfrequencies ($s_1$, $s_2$) are plotted in Figure 2a. As
we can see, the orbits of two planets precess retrogradely ($s_i<0$)
unless the gas disk is almost depleted ($\sigma_0\le 10 ~{\rm
gcm^{-2}}$), so the procession rate is slow (with period $> 2000$
yr) and prograde. Also the precession frequency $g$ located inside
2.0 AU dose not equal to  $s_1,s_2$ unless $\Sigma_g < 20~{\rm
gcm^{-2}}$ (Figure 2b-d). This occurs when $t> 3.8 $ Myr. At $t=10$
Myr, the gas disk is effectively depleted, either $s_1$ or $s_2$ may
meet $g$ at some locations of habitable zone. As the secular
resonances at habitable zone occurred only at the epoch when the gas
disk is almost depleted, we omit the effect of self gravity of gas
disk in the following simulations.

\vspace{4cm}

\section{Simulations and Results}

We numerically integrate the full equations of planet motions  with
a time-symmetric Hermit scheme (Aarseth 2003). Regularization
technique is  used to handle the collision between embryos, so all
the embryos have their physical radii. As long as the mutual
distance  between two embryos is smaller than the sum of their
physical radius, we assume they are merged.

We perform four Groups of simulations with different star accretion
rate ($\dot M$) and the heavy element scaling factor with respect to
the MMSN model ($f_d$), see Equation (\ref{miso}). In our
simulations we choose $\dot M=1\times 10^{-8},~5\times 10^{-8},~{\rm
and} ~2.5\times 10^{-7}~M_\odot / {\rm yr}$, corresponding to
different stages of T Tauri stars so that higher $\dot M$
corresponds to earlier stage.The conditions in different Groups are
listed as follows:
\begin{description}
\item[Group 1]:  $f_d = 2$; $\rm \dot {M}=5\times 10^{-8} M_{\odot}/yr$
responds to the green dot line in figure 1.
\item[Group 2]: $f_d = 2$; $\rm \dot {M}=1\times 10^{-8} \rm M_{\odot}/yr$
responds to the blue solid line in figure 1.
\item[Group 3]: $f_d = 2$; $\rm \dot {M}=2.5\times 10^{-7} \rm M_{\odot}/yr$
responds to the red dash line in figure 1.
\item[Group 4]: $f_d = 1$; $\rm \dot {M}=5\times 10^{-8} \rm M_{\odot}/yr$
responds to the green dot line in figure 1.
\end{description}
We perform 31 runs with the same disk model except varying $f_1$
from 0.001 to 1 in each Group. The results are presented as follows.

\subsection{Formation and orbital configurations of terrestrial planets}

One of the most important factors that affects the formation of
planets in habitable zone is the speed of type I migration reduced
by $f_1$. If the speed is very slow, the embryos  will not migrate
too much. Fig.3 shows a typical run of Group 1 with very low
migration speed ($f_1=0.004$). Planetary mergers mainly occur due to
scattering effect of embryo inside two giant plants at $t \sim 5 $
Myr (WZZ09). As we know from the density profile in Figure 1, when
$\rm \dot {M} = 5\times 10^{-8}\rm~ M_{\odot}/yr$, the location of
density maximum is at 0.3 AU.  At this density maximum, the type I
migration of embryos can be stalled due to the negative $\beta$ in
equations (\ref{tauI}). The inward migration of a second planet may
trap the stalled planet into its mean motion resonance, which might
increase the eccentricities of both planets. As we can see, most of
the planets survived at $t=10$ Myr with $a<1$ AU are trapped into
first order resonance (mainly 4:3). Among them, one is in the HZ and
two of them are at the edge of the HZ in this case.

With the increase of the type I migration speed, less and less
embryos are survived or trapped in HZ before the gas disk was almost
depleted. For example, Figure 4 shows a typical run with migration
reduced factor $f_1=0.08$. Only two terrestrial planets survived at
the inner region ($a< 1$ AU). Interestingly, they are in 2:1 MMR.
The inner planet with mass (7.5 $M_\oplus$) is in the HZ. As the
migration speed increases again, $f_1=0.25$, only one planet
survived in the HZ (Figure 5a). It is stalled under type I migration
at the inner edge of MRI dead zone (Figure 5b).

\begin{figure*}
 \epsscale{1}
\plotone{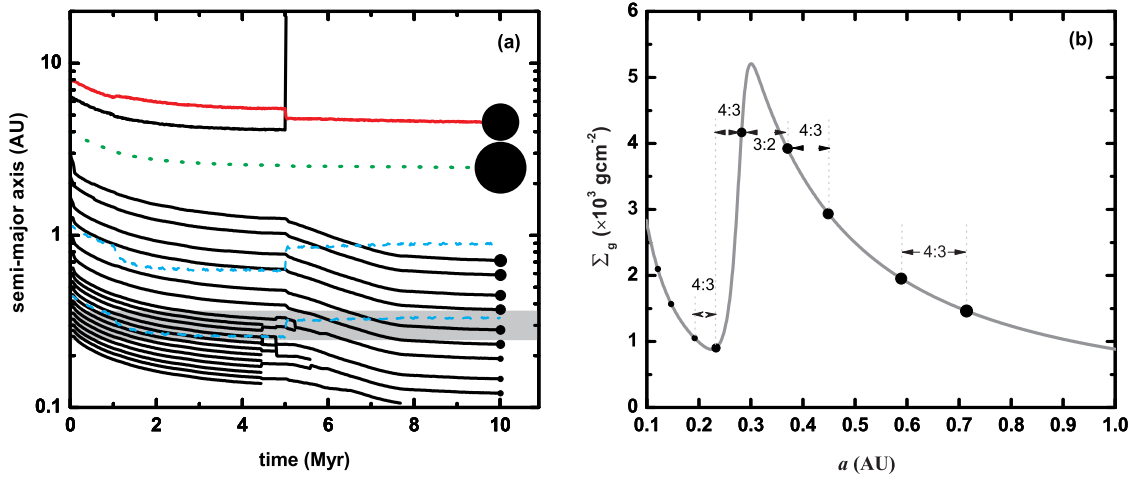}
 \vspace{-2cm}

 \caption{ A typical run with $f_1=0.004$ in Group 1. There are nine terrestrial
 planets with masses less than $7 M_\oplus$ left in the system. One is in
 the HZ and two of them are at the edge of the HZ.
 At the end of the simulation they trapped into 4:3 and 3:2 MMRs respectively. (a)
 Evolution of the semi-major axes. The gray band
 represents the HZ and the two blue dashed lines indicate the location of
 secular resonance caused by the giant planets.  (b) The final locations of the terrestrial planets in the
 system. The ordinate of each planet is chosen to fit the gas
 density profile for illustrative purposes. MMRs are labeled.
 \label{fig3}}
\end{figure*}

\begin{figure*}
 \epsscale{1}
\plotone{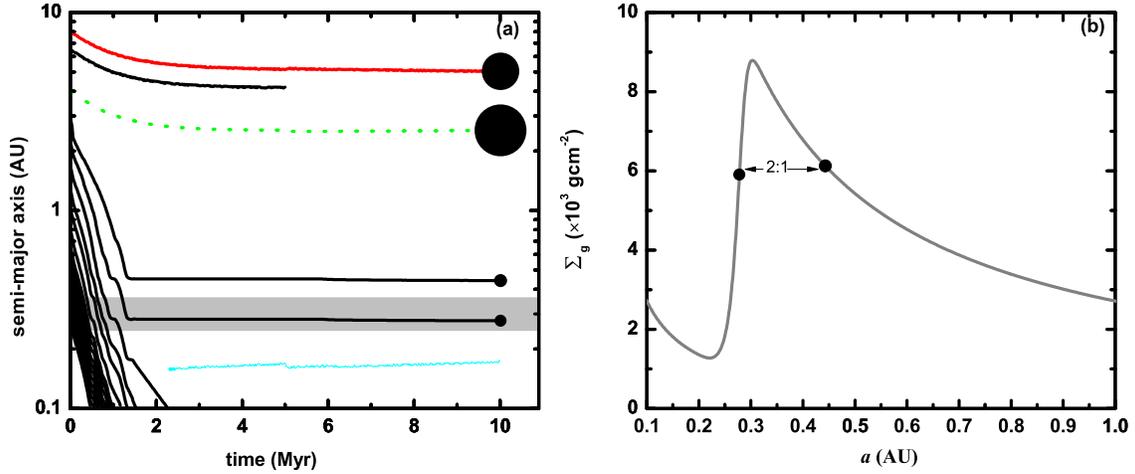}
 \vspace{-2cm}

 \caption{ A typical run with $f_1=0.08$ in Group1. There are two terrestrial
 planets left in the system. One is in
 the HZ. At the end of the simulation they trapped into 2:1. (a)
 Evolution of the semi-major axes. The gray band
 represents the HZ and the blue dashed line indicates the location of
 secular resonance caused by the two giant planets and the other terrestrial planet. (b) The final locations of the terrestrial planets in the
 system. The ordinate of each planet is chosen to fit the gas
 density profile for illustrative purposes.
 \label{fig4}}
\end{figure*}

\begin{figure*}
 \epsscale{1}
\plotone{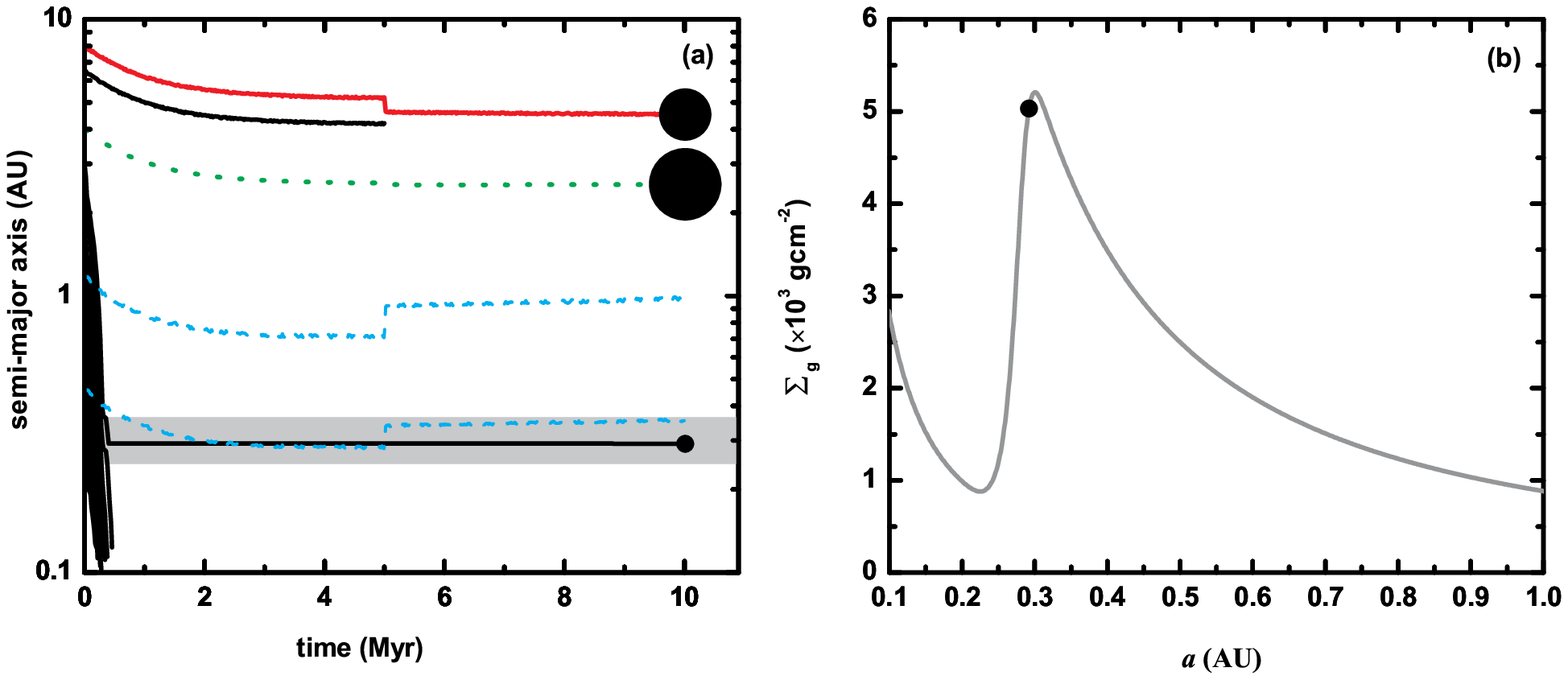}
 \vspace{-2cm}

 \caption{ A typical run with $f_1=0.25$ in Group1. There is one terrestrial
 planet left in the system and it is in
 the habitable zone. (a) Evolution of the semi-major axes. The gray band
 represents the HZ and the two blue dashed lines indication the location of
 secular resonance caused by the two giant planets.
 (b) The final location of the terrestrial planet in the
 system. The ordinate of the planet is chosen to fit the gas
 density profile for illustrative purposes.
 \label{fig5}}
\end{figure*}

According to Figure 3b,  many planet pairs may be trapped  in MMRs
under migration. Table 1 shows the statistics of the final
configurations for the simulations in the four Groups. We classify
the system being in either MMRs, aligned apsidal resonance (AARs) or
anti-aligned apsidal resonance (anti-AARs) if there is at least one
couple of planets trapped into MMRs, AARs or anti-AARs. Table 1
records the probability of a planet system that is in MMRs, AARs or
anti-AARs in the 31 runs of each group (24 runs in Group 3). As we
can see, more than a quarter of the systems contain at least a
planet pair that is in MMRs, AARs or anti-AARs.

\begin{table*}

\caption{Statistic of the final configurations of the simulations in
the four groups. f-Pb and f-Pc show the final average locations and
eccentricities of giant planet b and c respectively. f-MMRs is the
probabilities of planets (in all planets) and the system (with at
least one planet pair) captured into MMRs. We consider the system is
in MMRs if there is at least one couple of planets trapped into
MMRs. f-AAR and f-antiAAR are the statistic of the planets and
systems trapped into aligned apsidal resonance and antialigned
apsidal resonance in the final configurations. \vspace{0.5cm}
 \label{tbl-1}}
 \begin{center}
\begin{tabular*}{17cm}{@{\extracolsep{\fill}}llllll}
\tableline
  ID & f-Pb  & f-Pc &   f-MMRs   & f-AAR&  f-antiAAR  \\
              &\emph{a} (AU), e & \emph{{a}} (AU), e & Planet, System & Planet, System & Planet, System\\
\tableline
G1 & 2.474, 0.076 & 5.267, 0.108  & 63.4\%, 17/31   & 62.7\%, 18/31 & 56.2\%, 15/31 \\
G2 & 2.547, 0.087 & 5.272, 0.132  & 47.6\%, 16/31   & 48.4\%, 12/31 & 48.4\%, 16/31 \\
G3 & 2.486, 0.065 & 5.166, 0.101  & 30.8\%, 7/24    & 46.2\%, 6/24  & 27.4\%, 6/24  \\
G4 & 2.000, 0.087 & 5.040, 0.107  & 23.0\%, 11/31   & 56.0\%, 13/31 & 41.0\%, 13/31 \\
\tableline
\end{tabular*}
\end{center}

\end{table*}

\subsection{Stable time of the system}
For systems with only one terrestrial planet, we can get the stable time from
the results in WZZ09. Because the distance of the single terrestrial planet to the
 central star is closer than 0.56 AU in our results. All the single terrestrial planets systems
are stable in more than $10^8$ years.
 Based on Zhou et al. (2007) about the crossing time of planet system,
 we estimate the stable time of multiple terrestrial planet system at the end of the simulations.
We test each couple of the planets in the systems and use the average eccentricity,
 average mass and relative Hill radius for the two planets next to each other in the estimation.
Results are showed in table 2. More than 60\% couples of the planets in the system will survive
in more than $10^8$ years. According to the results of the stable time of the planets, we get the
stable time of the planetary system. We define the shortest time of the planet couples
surviving in the system as the stable time of the system.
 The results are also showed in the last column of table 2.
More than 41\% system will be stable in more than $10^8$ years.

\begin{table*}
\begin{center}
\caption{Statistic of the stable time of the multiple terrestrial planet systems. The first three columns means the statistic of the planet couples and the last column represents the results of the planetary systems in each group.
 \vspace{0.5cm}
 \label{tbl-2}}
\begin{tabular*}{12cm}{@{\extracolsep{\fill}}ccccc}
\tableline
  ID & $>10^6$ yr & $>10^7$ yr &   $>10^8$ yr & system stable time  $>10^8$ yr  \\
\tableline
G1 & 92.6\% & 84.4\%  & 63.9\% &41.67\% \\
G2 & 90.4\% & 86.2\%  & 69.1\%  &47.82\%\\
G3 & 91.1\% & 83.3\%  & 77.8\%  &50\%\\
G4 & 92.8\% & 92.8\%  & 87\%  &75.86\%  \\
\tableline
\end{tabular*}
\end{center}

\end{table*}

\subsection{Water contents of terrestrial planets}

For terrestrial planets in habitable zone, one important parameter
is the water content inside the rock. To model the distribution of
water contents within embryos, we initially assume the inner disk is
water-poor, embryo within 0.5 AU is 0.001\% by mass, while embryo
beyond 0.625 AU is wet with the water content 5\%, embryo between
them is moderately wet with 0.1\% water. We further assume, the
water content will be decreased by 30\% during each collision
(Melosh 2003). Thus the migration and merge of embryos may
re-distribute the water contents.

Figure 6 shows the simulation results of Group 1. The color map
represents the water distribution by mass (Raymond et al. 2004).
Panel (a1) and (b1) show the distribution of the habitable planets
changed with $f_1$. If type I migration of the embryos are fast,
$f_1> 0.2$, most of the small embryos originally in the HZ were run
into the central star, the only survival planet  with high mass
($\sim 7~M_\oplus$) was migrated from outside ($> 2.2$ AU). So they
have relative high water contents. When $f_1< 0.07$, the slow
migration of embryos are easily to be perturbed by either the
secular perturbations or secular resonance induced by the two
giants. So planetary merges occurred occasionally, and the motion of
the embryos are chaotic. As planetary merges reduce the water
contents of the rocks, such formed planets with masses $> 7
M_\oplus$ are drier than neighbor embryos .

\begin{figure*}
\begin{center}
  \epsscale{1.3} \plotone{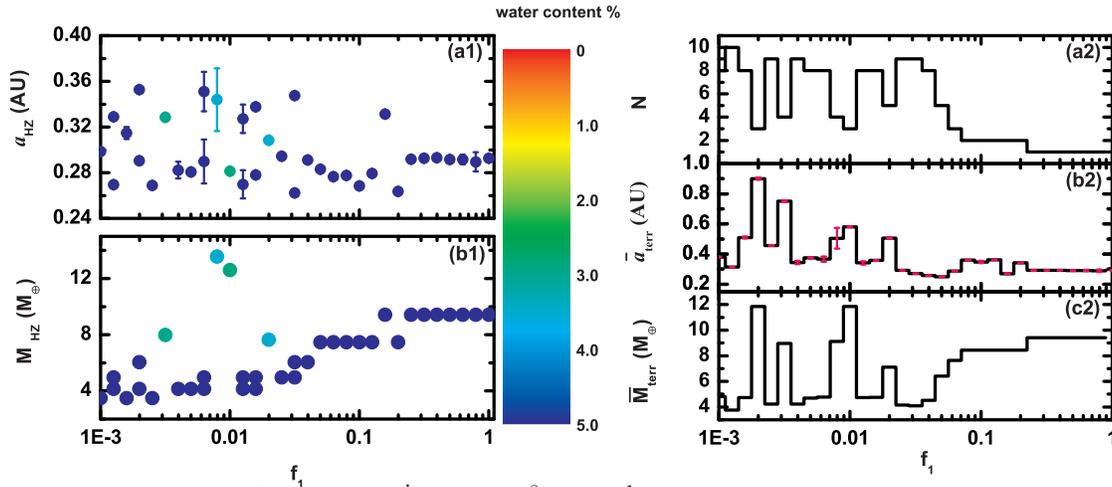}
 \vspace{-2cm}

 \caption{The results of Group 1 with $\dot M = 5\times 10^{-8}~M_\odot yr^{-1}$ at the end of our
simulations (t=10 Myr). (a1): The semimajor axes of the habitable
planets distribution changed with $f_1$. (b1): The Mass of the
habitable planets distribution. (a2): The number of the terrestrial
planets survived in the system. (b2) The average semimajor axes of
terrestrial planets distribution changed with $f_1$. (c2) The
average masses of the terrestrial planets distribution. The color
map in the left panels represents the water distribution by mass.
The error bars in Panel (a1) and (b2) show a(1-e) and a(1+e).
 \label{fig6}}
 \end{center}
\end{figure*}

Panels (a2), (b2), and (c2) of Figure 6 are the distribution of the
terrestrial planets. From Figure 6a2, if the type I migration speed
is slow, there will be many embryos left in the inner region. $f_1
\sim 0.1$ is the critical value $f_{1cri}$ above which only one or
two terrestrial planets formed and survived in the system. The
average semi-major axes of the terrestrial planets showed in Panel
(b2) are small when $f_1 \ge 0.1$, within the snowline (0.68 AU in
OGLE-06-109L system), so most of the terrestrial planets keep in the
inner region of the system. As it's possible to form planets with
mass $>10~M_\oplus$, they might have the chance to accretion gas and
form Neptune-size planets if there is still enough gas in the disk.

In Group 2 and 3, we change the star accretion rate to $\rm \dot
{M}=1\times 10^{-8}~ \rm M_{\odot}/yr$ and $\rm \dot {M}=2.5\times
10^{-7} ~\rm M_{\odot}/yr$ respectively. This corresponds to the
formation of terrestrial planet occurred either later or earlier so
that the mass accretion rate of the star is either smaller or
larger. Figure 7 shows the results of the two groups. When $f_1>
0.1$ of Group 2, there are one or two planets left in the system,
and  the inner most terrestrial planet with mass $\sim 9.4~M_\oplus$
locates around 0.22 AU, the density profile maximum around the HZ.
For Group 3, $\dot M =2.5\times 10^{-7}\rm M_{\odot}/yr$, the
location of density maximum is 0.62 AU. There is no terrestrial
planet left in the system for $f_1> 0.2$.

From Figure 7c1 and c2,  we can get  that if $f_1 \le 0.1$, there
are more than one terrestrial planets survival in the system. It
also agrees with the result of Terquem \& Papaloizou (2007) that the
hot super-Earths or Neptunes formed by mergers of inwardly migrating
cores are most likely not isolated. The average semimajor axes in
Figure 7d1 and d2 show that most of the terrestrial planets are
distributed inside the snowline (0.68 AU). The water content in the
region of $f_1 < 0.1$ is lower than that of $f_1 > 0.1$,  in which
case the terrestrial planets are mainly from outside orbits with
high water contents.

\begin{figure*}
\begin{center}
 \epsscale{1.30} \plotone{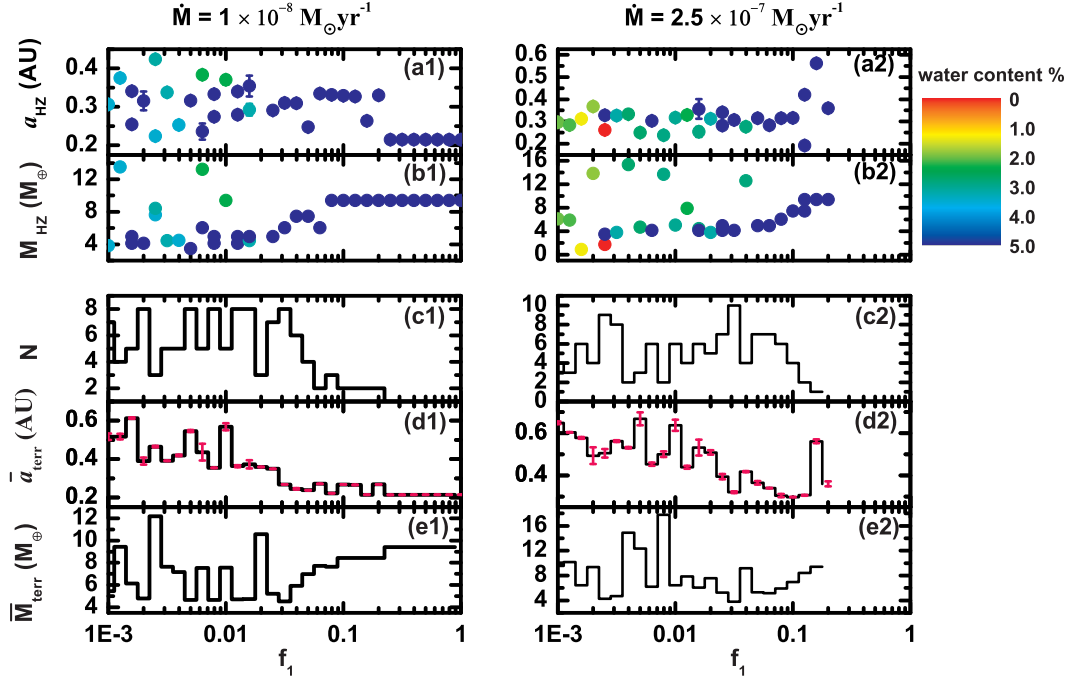}
 \vspace{-0.5cm}

 \caption{ The results of Group 2 (left panels )  and Group 3 (right panels)
  at the end of our simulations (t=10 Myr). Panel (a1) and (b1) are the
 semimajor axes and mass distribution of the habitable planets in
 Group 2 with $\dot M = 1\times 10^{-8}~M_\odot yr^{-1}$.
 Panel (c1), (d1), and (e1) are the number, average
 semimajor axes, and average mass distribution of the terrestrial
 planets in Group 2. Panel (a2) and (b2) are the results of habitable planets
 in Group 3 with $\dot M = 2.5\times 10^{-7}~M_\odot yr^{-1}$. Panel (c2), (d2), and
 (e2) are the results of terrestrial planets of Group 3.
 The error bars in panel (a1), (a2), (d1), and (d2) show a(1-e) and a(1+e).
 \label{fig7}}
 \end{center}
\end{figure*}

Figure 8 shows the results of Group 4 for the embryos with $f_d =
1$. Because $f_d$ is related to the mass of the embryos initially,
the mass of the habitable planets will be smaller than other Groups.
From the timescale of type I migration, equation (6), the migration
speed is in proportion to $f_1\times m $. So when $f_d = 1$, the
migration is slower than the runs in other Groups. In Group 4, if
$f_1$ is less than 0.2, most of the habitable planets left in the
system are merged from small embryos for several times. The water
contents of the planets are low. When $f_d = 1$, the initial masses
of all embryos are less than 3.3 $M_{\oplus}$. At last the average
mass of Group 4 is about 4 $M_{\oplus}$. The habitable planet whose
mass higher than 7 $M_\oplus$ is formed from merge with low water
content. In this Group, there are more than one terrestrial planets
survival in the range of $f_1\in [0.001,1]$.

\begin{figure*}
 \epsscale{1.30}
\plotone{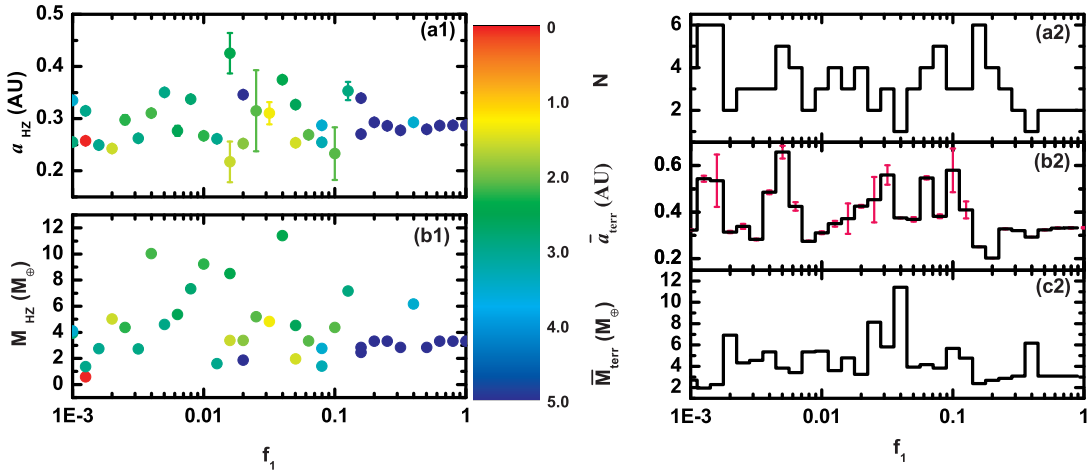}
 \vspace{-1cm}

 \caption{ The results of Group 4 at the end of our simulations (t=10 Myr). Panel (a1) and (b1) are the
 habitable planets distribution of Group 4 with fd = 1.
 Panel (a2), (b2), and (c2) are the results of terrestrial planets in
Group 4. The error bars in Panel (a1) and (b2) show a(1-e) and
a(1+e).
 \label{fig8}}
\end{figure*}

\section{Conclusions and Discussions}

In this paper, we investigate the formation of terrestrial planets
under the perturbation of giant planets around dwarf stars, with a
model system OGLE-06-109L. We assume two giant planets already
formed in outside orbits. The embryos are initially embedded in an
exponentially decaying gas disk and will under type I migration with
a speed reducing factor $f_1$ over the standard (linear) estimation.
The presence of giant planets may excite the eccentricities of
embryos in nearby orbits through secular perturbations, and induce
secular resonance in the habitable zone regions when the gas disk is
almost depleted.

The speed of type I migration plays an important role in determine
the number and orbital architecture of terrestrial planets. When
$f_1\le 0.1$ (0.1 times of the standard speed), there may be several
terrestrial planets formed in the inner region ($<1 $ AU). Massive
terrestrial planets formed by mergers of smaller mass embryos, thus
they have lower water contents compared with neighboring embryos
with small masses. Some of the planet pairs are trapped either in
MMRs, or AARs or anti-AARs. The formation of planets in MMR is due
to the halt of inner planets at the inner edge of MRI dead zone, or
due to the convergent migration during the evolution. According to
our simulations, most of the formed MMRs are first order resonances
which have wider resonance regions. In the case that type I
migration speed is fast ($f_1>0.1$), there will be very few (or
sometimes no) terrestrial planets in the inner region. They are
mainly migrated from outer regions so they have higher water
contents as compared with neighboring embryos. Thus through the
water contents of the terrestrial planets, we can deduce their
origin.

The change of star accretion rate reflects the location of the
density maximum, which affects the survival of the planets in the
habitable region [0.25, 0.36] AU in OGLE-06-109L system. If the
giant embryos in the inner region ($<$ 1AU) of the system formed in
the earlier (for example, Figure 7a2-e2, $\dot M=2.5 \times 10^{-7}~
M_\odot/yr$), the surface density of the disk is still high so that
the high speed type I migration (with $f_1>0.1$) may drive all the
embryos into the star. Only  lower speed migration and later formed
embryos are easier to retained. However, the model of type I
migration described by equations (\ref{tauI}) and (\ref {FI}) are
linear. When more realized  model is considered, e.g., considering
the outward migration in  viscous, radiative disks (Kley et al.
2009), things will be more complicated and interesting.

Another important factor is the initial conditions of the
embryos. In our simulations, the embryos we used are in isolation
masses. In order to test the reasonability of the initial
conditions, we run with more embryos with smaller initial masses
using $f_1=0.004$. Figure 9 shows the result. Finally, we get a
planet with mass of 4.4 M$_{\oplus}$ in the habitable zone which
is comparable to the result in figure 3. And the system will be
stable in more than $10^8$ years. We test other runs and find that
the results are similar to the run when $f_1=0.004$. So our
results with the embryos in isolation mass initially are
reasonable.

\begin{figure}
 \epsscale{1.2}
\plotone{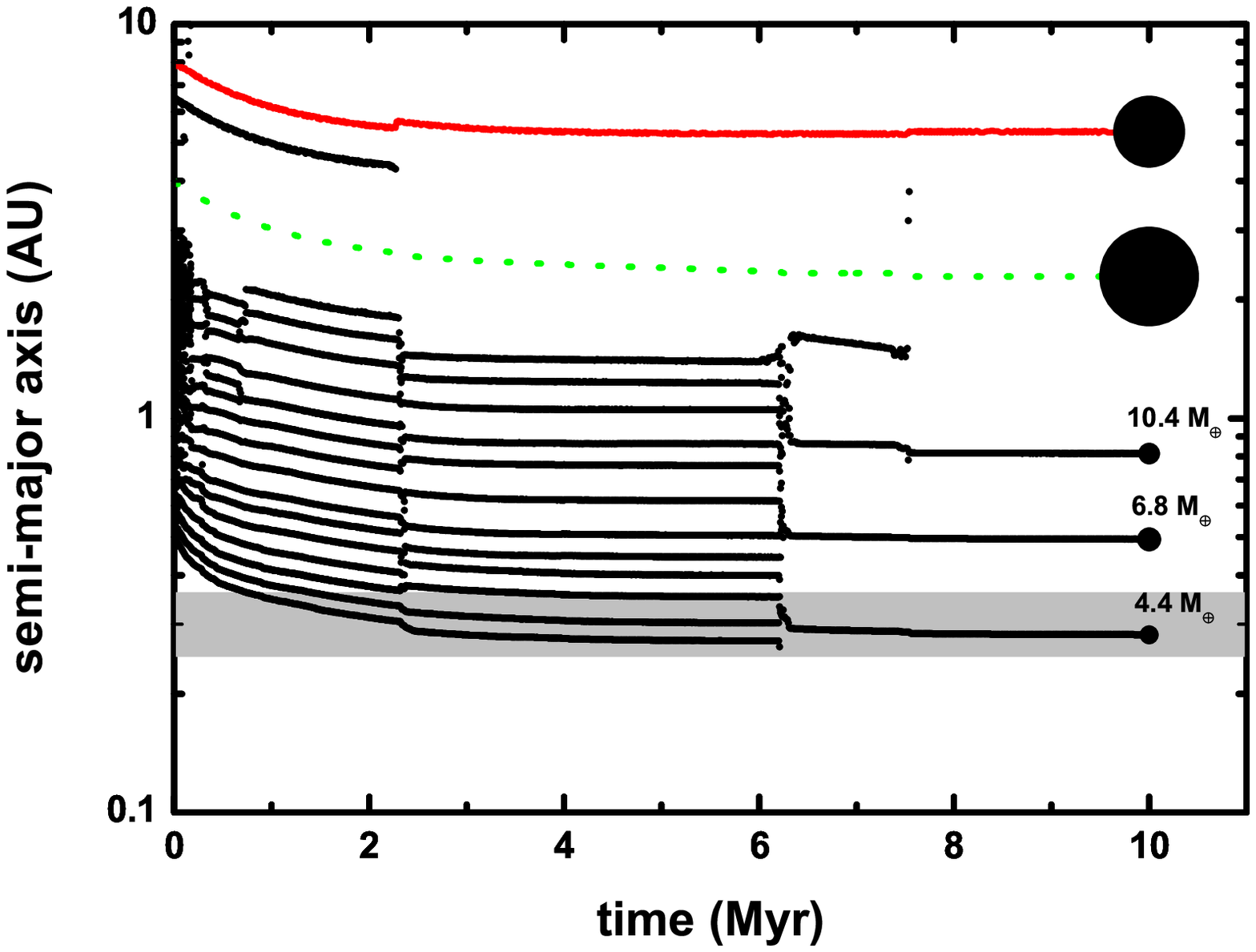}
 \vspace{0cm}

 \caption{ The results with different initial conditions.
 In this run, we use 90 embryos in 0.4 M$_{\oplus}$ initially.
 Finally, we get a planet with mass of 4.4 M$_{\oplus}$ in the habitable zone which is comparable to the result in Figure 3.
 \label{fig9}}
\end{figure}

According to our simulations, by checking the water contents of
terrestrial planets in OGLE-06-109L (and other similar systems), one
can understand their formation history, and also restrict the
parameters such as $f_1$ used in the planetary formation theory. For
the OGLE-06-109L system, the snowline is about 0.68 AU, very close
to the central star.  If one or no terrestrial planet is observed,
basically they might formed through inward migration from outside
the snow line, thus they will have relatively high water contents
($\sim 0.5\%$), also the type I migration speed is fast ($f_1>0.1$).
If many planets are observed in the inner region ($ <1$ AU), then
most probably, they are formed through merges of isolation cores,
thus they will have relatively low water contents $\sim 2\% $. The
presence of multiple terrestrial planets also indicates a low speed
migration of the embryos ($f_1<0.1)$.

This work is supported by NSFC (10925313, 10833001, 10778603),
National Basic Research Program of China (2007CB814800) and Doctoral Funds of Universities (20000091110002).

\end{document}